\documentclass[submission,copyright]{eptcs} % change for EPTCS submission
\providecommand{\event}{ACL2 2011}
\usepackage{breakurl}
\usepackage{epsfig}

\title{Toward the Verification of a Simple Hypervisor}

\author{Mike Dahlin, Ryan Johnson, Robert Bellarmine Krug, Michael McCoyd, William Young
\institute{Department of Computer Sciences\\
University of Texas at Austin}
\email{\{dahlin, rjohnson, rkrug, mccoyd, byoung\}@cs.utexas.edu}
}

\def\titlerunning{Verification of a Simple Hypervisor}
\def\authorrunning{M. Dahlin, R. Johnson, R.B. Krug, M. McCoyd, W. Young}

\date{4 July 2010}

\begin{document}
\maketitle

\begin{abstract}
Virtualization promises significant benefits in security, efficiency,
dependability, and cost.  Achieving these benefits depends upon
the reliability of the underlying virtual machine monitors
(hypervisors).  This paper describes an ongoing project to develop and
verify MinVisor, a simple but functional Type-I x86 hypervisor,
proving protection properties at the assembly level using ACL2.
Originally based on an existing research hypervisor, MinVisor provides
protection of its own memory from a malicious guest.  Our long-term
goal is to fully verify MinVisor, providing a vehicle to
investigate the modeling and verification of hypervisors at the
implementation level, and also a basis for further systems research.
Functional segments of the MinVisor C code base are translated into
Y86 assembly, and verified with respect to the Y86 model.  The
inductive assertions (also known as ``compositional cutpoints'')
methodology is used to prove the correctness of the code. The proof of
the code that sets up the nested page tables is described.  We compare
this project to related efforts in systems code verification and
outline some useful steps forward.
\end{abstract}

%A category including the fourth, optional field follows...
%\category{D.2.8}{Software Engineering}{Metrics}[complexity measures, performance measures]

%\terms{Virtualization, Hypervisors, Formal Verification, ACL2, Theorem Proving}

%=============================================
\section{Introduction}
\label{Introduction}

Platform virtualization refers to technologies that provide a layer of
abstraction between computer systems and the operating systems that
utilize them.  It has existed for many years and appeared in various
guises, but has recently emerged as a key technology with vast
potential impact on commercial and government computing
systems. Platform virtualization promises significant benefits in
security, efficiency, dependability and cost~\cite{HessNewman}.
Achieving these benefits depends upon the reliability of the virtual
machine monitors (hypervisors) that provide them.

We believe that providing high assurance of the properties of a system
as complex as a hypervisor requires a rigorous formal analysis,
stretching from the desired security properties at the top, down to
the physical hardware and implementation code at the bottom.  This
project aims to develop a simple but useful hypervisor and to provide
very high assurance in its properties by using ACL2~\cite{ACL2Book}
to verify the hypervisor implementation.

For the purposes of this paper, a \textit{hypervisor}, also called a
virtual machine monitor (VMM), is a software system that virtualizes
some system resources of the host computer and makes them available to
one or more guests.  Hypervisors provide a range of
functionalities~\cite{SmithNair} and a diversity of guarantees.
We chose to pursue verification of a hypervisor because of the
increasing impact of hypervisors on commercial and government systems.
However, existing commodity hypervisors have a large, complex and
changing code base and seem beyond the reach of current modeling and
verification techniques.  Instead, we focused on the verification of a
small research hypervisor that we could envision tackling with the
limited resources available to the project, in hopes that the tools
and techniques developed might scale to larger system software
verification efforts.  

Though our target hypervisor, called MinVisor, runs only one guest, its
features provide some of the minimum requirements of typical
hypervisors and provide some of the guarantees of CMU's
SecVisor~\cite{SecVisor}, a simple research hypervisor that we used as
our initial jumping off point.  Thus, MinVisor provides a simplified
but useful platform on which to carry out our initial verification
efforts.

This paper reports on the development of MinVisor and progress
in its modeling and verification.  Though early, the project has had
some significant success in moving toward using formal verification to
develop a simple hypervisor with very high assurance. 

In Section \ref{High Level Motivation} we describe the overall
strategy of the project.  Section \ref{MinVisor} outlines the
hypervisor artifact we are constructing.  Section \ref{Y86++}
describes the Y86++ model on which our proofs are constructed.  In
Section \ref{Implementation Level Proofs} we outline our proof
strategy and describe the proof of specific critical MinVisor
functionality.  Section \ref{Related Work} lists some relevant related
work.  Finally, in Section \ref{Conclusions and Future Work} we
outline our plans for extending this effort.  The Appendix contains
the C source code and Y86++ assembly level code for specific MinVisor
functions.

\section{High Level Strategy}
\label{High Level Motivation}

The overall goal of this project is to investigate techniques for
modeling and verifying a simple hypervisor.  At the end of the day, we
hope to demonstrate that:
\begin{enumerate}
\item We can carry out the verification of a hypervisor at a very low
  level of abstraction to provide high assurance of correctness.
\item The resulting verified artifact is a credible piece of
  system software displaying useful hypervisor functionality.
\item The tools and techniques we develop will scale to apply to
  more realistic and full featured hypervisors. 
\end{enumerate}

Modeling and verification of a complex artifact might follow a
top-down approach---taking an abstract model of the domain
(hypervisors in this case) and refining it toward a compliant
implementation.  Alternatively, it might follow a bottom-up
approach---extracting and proving properties of implementation level
modules with the goal of combining these into a coherent verified
model of the complete artifact.  A hybrid approach is probably best,
to ensure that the two ``meet in the middle.''

To date, we have worked largely bottom up, concentrating on developing
the techniques necessary for verifying the many lines of machine code
that comprise the MinVisor code base.  We felt that defining a
reasonable machine model and building the libraries and infrastructure
to allow efficiently proving significant fragments of code was
probably our biggest hurdle.  Consequently, the description of the
modeling and verification below is focused on the low level.  We
describe the verification of the code that sets up MinVisor's nested
page tables.  This functionality is critical to allow MinVisor to
protect itself from a malicious guest, and seems representative of the
code complexity we will encounter moving forward.  We believe that
this or very similar code will be required for any hypervisor.

Our initial code base was extracted from CMU's SecVisor's
code~\cite{SecVisor}.  However, we made various simplifications to
eliminate functionality orthogonal to our concerns and to simplify our
initial proof target.  For example, SecVisor is tightly coupled with
its modified Linux guest, making Linux part of the trusted computing
base.  By making various modifications we were able to eliminate this
dependency, yielding a simpler code base to verify and a more broadly
applicable hypervisor.  MinVisor loads the guest as if by the BIOS and
protects the hypervisor state via nested paging.

Protection of the core system and software is crucial to any system
making security claims, and is the primary initial goal of this
project. However, in order to be useful, hypervisors must also provide
a reasonable operating environment to the guest.  Our current work
does not perform any sort of corrective emulation, nor any
paravirtualization to provide alternate resource access.
The unmodified guest runs freely as long as it
performs only ``safe'' actions.  Dangerous operations are intercepted
and prevented, but MinVisor's operations when doing so may be
detectable by the guest.  Providing a reasonably transparent and accurate
simulation of a physical machine is future work.

\section{MinVisor}
% please do not fill (word wrap) paragraphs in this section
\label{MinVisor}
% other places w/ MinVisor descriptions: Raytheon 2011-Q1 sec 2.1

MinVisor was developed to examine the proof of the protection of a
hypervisor and machine from a malicious guest and as a base for
further systems work.  It is a minimal Type-I/bare-metal hypervisor
for the x86 utilizing AMD's hardware assisted AMD-V nested page table
virtualization.  The hardware BIOS loads MinVisor from the network
using the standard BIOS PreBoot Execution Environment
(PXE)~\cite{PXE}.  MinVisor loads the first hard drive sector, as if
by the BIOS, as the guest.  On the QEMU x86 emulator it runs a real
and protected mode guest that tries to probe memory in each 2MiB
machine page.  It has not been tested on real hardware.  MinVisor is
written in a mix of assembly and C, limited to 32 bit instructions,
and draws from code in SecVisor, JOS~\cite{JOS}, and Xen~\cite{XEN}.

MinVisor was designed as a real hardware, tiny, hypervisor
that does the essential tasks of any AMD-V nested paging 
 hypervisor as simply as possible.
These essential tasks are:
\begin{itemize}
\item Create host save area and VMCB,
\item Create IO and MSR permission bit maps,
\item Set guest actions to intercept,
\item Populate nested page tables,
\item Set starting guest state in VMCB, 
\item Load guest code, 
\item Populate and activate host page tables, 
\item Start the guest, and
\item Handle intercepts (for nested page faults).
\end{itemize}
Except for status messages and memory read values (below)
MinVisor currently does only these things.

% History relation to SecVisor 

Though MinVisor started from SecVisor, we have made many
simplifications and the code is now closer to KVM~\cite{KVM} in
organization.  We have focused on providing absolute protection of the
host and hardware from an arbitrary and potentially malicious guest.
Secondarily, we provide some transparency to the guest by booting over
PXE which leaves no hard drive evidence of our existence.  Our initial
phase aims to protect MinVisor's memory.  If MinVisor's memory is not
protected, no guarantees can be made about its operation. 

MinVisor is the first code to run after the BIOS, whose PXE tftp code
downloads MinVisor to physical address 0x7C00.  Control is transferred
to MinVisor in real mode, and MinVisor makes no further use of the
BIOS PXE tftp capabilities.  After relocating itself to the highest available
2MiB aligned memory chunk and establishing protections, MinVisor emulates the behavior of
the BIOS had PXE not been called, and copies the first sector from the
hard disk to 0x7C00, acting as a real mode boot loader.  We assume a
static root of trust of the boot path.  The guest is currently limited
to 4GiB of RAM.

MinVisor emulates a normal system startup with its real mode start of
the hard drive's boot sector as the guest.  This minimizes
interdependencies between MinVisor and the guest.  MinVisor need not
receive any information from the guest, and the guest need not be
aware that MinVisor is running.  It should allow any hard drive
bootable guest to be run, and it allows MinVisor to run on any machine
with PXE, which is widespread in both consumer and enterprise
hardware.  Booting MinVisor with PXE allows providing the entire hard
drive to the guest, avoiding the need to virtualize it.

Most control over the guest is exercised through proper configuration
of the virtual machine control block (VMCB), a per-guest memory region
maintained by the hypervisor and virtualization hardware that contains
settings and information necessary to launch a guest
through AMD-V mechanisms.  The VMCB points to the nested page table
that provides protection from direct access to MinVisor's own memory.

% Paging background

Normally, process effective addresses are translated to physical
addresses through segmentation or segmentation with paging.  First, the
addresses are combined with the appropriate segment register to create
a linear address.  If page tables are used, this linear (or virtual)
address then serves as the index into the page tables and a physical
address is obtained.  It is this physical address that is used by the
hardware.

% Nested background

However, a third translation is involved for guest addresses if
hardware virtualization is using \textit{nested} paging. The guest
physical address obtained by the first lookup is used to index into
the nested page tables, and a true system physical address is
obtained.  These nested page tables are used by MinVisor to protect
its memory regions, and are more fully described in Sec. \ref{Changes
  from Y86}.

% MinVisor's mapping

Except for a few regions, MinVisor's nested page tables establish an
identity map from guest ``physical'' to system physical memory; most
addresses in the guest will resolve to the same location as they would
if the guest were running on the bare machine.  The identity map was
chosen for simplicity, though a more complex mapping would have been
only marginally more difficult to implement and verify.  The exception
to the mapping is that addresses in the top 2MiB of physical memory
have an empty ``not present'' mapping from the guest.  Thus, no guest
address will be translated to the physical memory where MinVisor is
located.

Should the guest, through ignorance or malice, try to access
an address marked ``not present'' in the nested page tables, the AMD-V
hardware will generate an intercept and return to host mode with
information about the guest action.  In such cases, the MinVisor
intercept handler causes writes to protected regions to be no-ops, and
reads return bytes from within the 16 byte string ``Th Eyes of
Texas''.  Use of the hardware architecture's nested page tables
greatly simplifies MinVisor and its proof, and allows the guest to run
uninterrupted most of the time. To help the guest use reliable memory,
we virtualize the BIOS memory map to report the top of physical memory
as 2MiB lower than it actually is.

Interception of guest access to hardware I/O is restricted via a table
referenced in the VMCB. We currently do not intercept any hardware
ports, but plan to default to intercepting all of them. This
interception requires some emulation of functionality in the intercept
routines in order to provide required I/O services to the guest.

The VMCB also contains a bit map allowing selective interception of
various operations. We currently intercept all operations related to
interrupts, control registers, and virtualization and convert them to
no-ops. In order to provide a reasonable operating environment to a
non-trivial guest, we will need to emulate or forward basic interrupt
handling and some system control events.

Initially we are implementing paranoid protection from DMA enabled
devices. Rather than intelligently map and cordon devices, all devices
are prevented from writing to the location of MinVisor or to memory
mapped I/O regions. This satisfies our initial protection and code simplicity
goals, with performance and functionality issues to be addressed
later.

Though MinVisor is sparsely featured by combination of design and
its early stage of development, we feel it is still a reasonable hypervisor
target for a proof of implementation. The core memory protection mechanisms
it provides by way of hardware initialization and software routines
are necessary functionality for any hypervisor. Additional features and
functions will build on top of the initial code and proof, rather
than supplant it.

\section{The Y86++ Model}
\label{The Y86++ Model}
\label{Y86++}

The goal of this project is a simple hypervisor verified at the
implementation level.  This requires a formal model of the
implementation language, and the machine it runs on.  Several related
efforts~\cite{seL4, Nova} choose to formalize subsets of C/C++.  We
opted instead to conduct our proofs at the machine code level.  This
choice eliminates relying on any formal model of C/C++ and capitalizes
on strengths of ACL2.  But ideally, it means verifying code against a
formalization in ACL2 of the x86 instruction set architecture.  Though
colleagues are working on such a model, it has not reached a stable
and usable state.

To allow progress on this project, we began our proofs instead on a
modified version of Bryant and O'Hallaron's Y86
model~\cite{BryantOHallaron}.  The Y86, designed for pedagogical
purposes, is a much simplified version of the x86 instruction set
architecture.  The Y86 was originally formalized in ACL2 by Sandip
Ray~\cite{Raydissertation} for his dissertation work.  Another
formalization in ACL2 was developed by Warren Hunt for use in his
computer architecture classes at the University of Texas at Austin.

Over the course of this project, we have modified Hunt's Y86 model by
adding several components to the processor state, several new
instructions, a guest/supervisor mode flag, and nested page tables.
In this paper, we refer to our extension of the Y86 model as the
Y86++, to prevent confusion with either Bryant and O'Hallaron's Y86 or
Hunt's implementation of it in ACL2.  Because of the similarity of the
Y86++ to the x86, we believe that the verification work we have done
to date will translate rather directly to an x86 model when one
becomes available.  The proof techniques and invariants are generally
independent of the details of the Y86++.

\subsection{Changes from Y86}
\label{Changes from Y86}

\begin{figure}
\begin{center}
\includegraphics[scale=0.6]{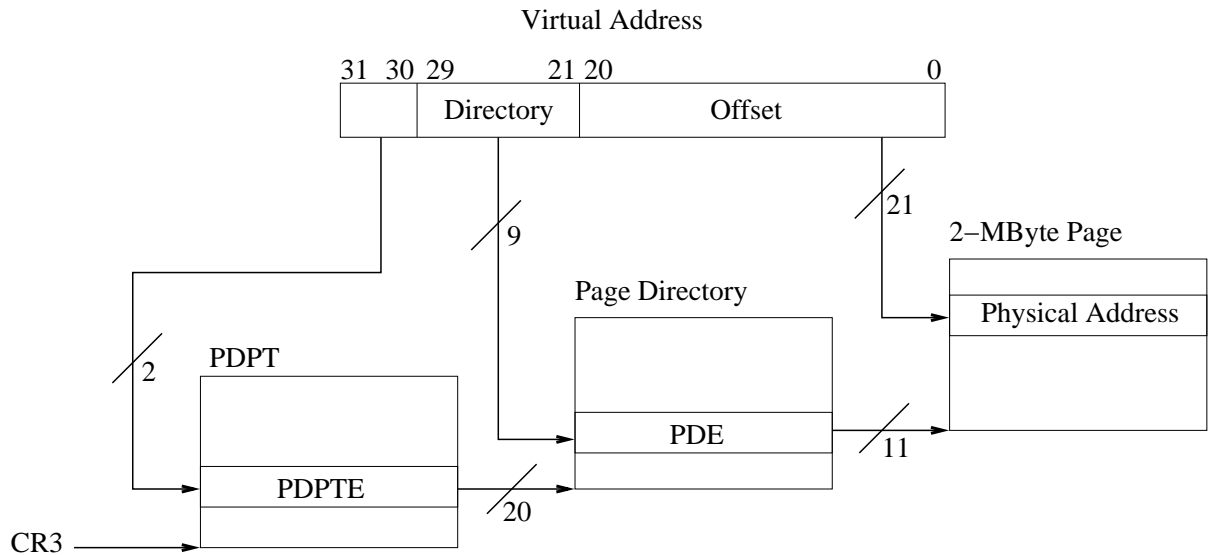}
\end{center}

\caption{\label{address-translation} Linear Address Translation to a 2-MiByte Page}
\end{figure}

The Y86 is an instruction set architecture in the style of
the x86.  We refer the interested reader to~\cite{BryantOHallaron} for
details.  We have made the following additions to Bryant
and O'Hallaron's Y86:

\begin{itemize}
\item Two new registers {\tt IMME1}, {\tt VALU1}: used as ``scratch
  registers'' to more closely emulate specific complex x86
  instructions as a series of Y86++ instructions.  Examples of their
  use will be given in Section \ref{Comparison to the x86}.

\item A register {\tt CR3}: points to the base of the page table
  hierarchy.

\item A carry flag:  used in several of the new instructions. 

\item A 1-bit Guest/Supervisor Mode flag, and four 32-bit registers:
  these allow mode switching and preservation of {\tt ESP} and {\tt
    EIP} when switching modes.
\end{itemize}
In addition, we have added the following instructions: 
\begin{itemize}
\item {\tt adcl} (add with carry), 
\item {\tt cmpl} (compare), 
\item {\tt sall} (arithmetic shift left),
\item {\tt shrl} (shift right), 
\item {\tt jb} (jump if below), 
\item {\tt jbe} (jump if below or equal).  
\end{itemize}

Memory in the Y86++, as in Hunt's implementation of the Y86, is
modeled as an association list from 32-bit addresses to 8-bit values.
We have implemented a nested page table mechanism on top of that.
Paging is enabled only when the Y86++ is in Guest Mode.

The Y86++ paging mode is modeled on Intel's PAE paging, which is what
is used by SecVisor and MinVisor for the nested page tables.  With PAE
paging (see Figure \ref{address-translation}), the first paging
structure comprises only $4$ entries and is pointed to by the {\tt
  CR3} register. Translation uses bits $31:30$ from a 32-bit virtual
linear address to index the appropriate entry. This entry points to
the appropriate second level paging structure, comprising $512 =2^9$
entries.  Bits $29:21$ identify the appropriate entry pointing to a
2MiB page frame.  Both entries also contain certain bits identifying
the status of the page frame.  The only status bit we concern
ourselves with at the moment is the \texttt{page present} bit.  If this bit
is 0 (as is the case for all pages that MinVisor protects), a page
fault is thrown if the page is accessed.  
At the moment, the Y86++ does not properly handle
page faults. This is not relevant to the theorems we have proved, but
will be relevant in subsequent work.

We have implemented an assembler in ACL2 to convert Y86++ assembly
language programs into their binary format, based on Hunt's Y86
assembler.  This is merely a convenience.  Our proof is at the level
of the binary;  its validity does not depend on the assembler.

\subsection{Comparison to the x86}
\label{Comparison to the x86}

The Y86++ assembly is similar to a restricted subset of the x86.
Other than the limited number of instructions modeled, the main
restriction is in the handling of immediate values --- the only
instruction that handles these is the {\tt irmovl} (immediate to
register move) instruction.  

The following is a fragment of Y86++ assembly generated by hand from
the corresponding x86 instructions (in the comment column).  The x86
assembly was generated by the gcc compiler from a portion of the
MinVisor implementation.
\begin{verbatim}
:init_pdts                 ;  init_pdts:
(pushl :ebp)               ;  pushl %ebp
(rrmovl :esp :ebp)         ;  movl  %esp, %ebp
(pushl :esi)               ;  pushl %esi
(pushl :ebx)               ;  pushl %ebx
(irmovl 48 :imme1)         ;  subl  $48, %esp
(subl :imme1 :esp)         
(irmovl 231 :imme1)        ;  movl  $231, -24(%ebp)
(rmmovl :imme1 -24 (:ebp))
(irmovl 1 :imme1)          ;  addl  $1, -28(%ebp)
(mrmovl -28 (:ebp) :valu1)
(addl :imme1 :valu1)
(rmmovl :valu1 -28 (:ebp))
\end{verbatim}

Note that many of the statements translate directly.  The more
complex translations involve instruction modes (using immediate values)
not present in the Y86 or Y86++ models.  For example, subtracting an
immediate value from a memory location (e.g., {\tt sub1 \$48, \%esp})
requires moving the immediate value into a register first, since only
immediate to register operations are modeled.

In each case, the translation has been crafted to match the semantics
of the corresponding x86 instruction.  The addition of the {\tt
  :imme1} and {\tt :valu1} registers makes this translation both
simple and direct.  We believe that the resulting state of the Y86++
memory and the state of all registers and flags that correspond to x86
structures are identical to what follows the execution of the original
x86 instructions.  That claim has not been formally established,
however, and cannot be until we have an x86 model in hand.  When we
do, we plan to switch our proof effort to the x86 model, and this
question will become moot.

\section{Implementation Level Proofs}
%\section{Implementation Level Proofs}
\label{Implementation Level Proofs}

MinVisor code is verified following the ``cutpoints'' approach
described elsewhere~\cite{ACL2cutpoints,Moore-cutpoints}.  Given an operational
semantics for a machine language and an annotated program, this allows
the mechanized generation of verification conditions adequate to
establish the partial (or total) correctness of the code.  
ACL2 is then used to dispatch the verification conditions.  

The method is compositional, in that the correctness of a subroutine
needs to be proved once, rather than at each call site. The method has
been used to verify several machine-level programs prior to the
current project using the ACL2 theorem prover.

To date, we have completed proofs of selected portions of MinVisor
codebase.  In particular, we describe below a proof of the functions
that set up the MinVisor nested page tables.  Those pages that
MinVisor wishes to protect---its own memory image and the nested page
table structures---are marked as \texttt{not present}.  Address
translation for the rest of memory is the identity map.  The process
of setting up the page tables is described in the following
subsection. Then in Section \ref{A Sample Theorem} we give a
description of the ACL2 modeling and proof.

\subsection{Correctness of the Page Table Code}
\label{Correctness of the Page Tables}
\label{Setting Up Page Tables}

In our Y86++ model, the function \texttt{(va-to-pa addr s)} is
responsible for the translation from a virtual address \texttt{addr} to
a physical address via a page table lookup in the state \texttt{s}.  If
paging is turned off, \texttt{va-to-pa} reduces to the identity map on
\texttt{addr}.  If paging is turned on, address translation is done via
the page tables as indicated in Figure \ref{address-translation}.
Normally, this will result in \texttt{va-to-pa} returning an address.
But, if during this lookup the entries used contains a \texttt{page
present} bit set to 0, a page fault is raised.

The C function \texttt{create\_nested\_pt} (actually the corresponding
Y86++ binary) is responsible for setting up the nested page tables to
be used by MinVisor.  The appendix contains the C source code and a
sample of the corresponding Y86++ assembly level code.
\texttt{create\_nested\_pt} calls three subsidiary functions:
\texttt{init\_pdpt} sets up the top level Page Directory Pointer
Table; \texttt{init\_pdts} sets up the four second level Page
Directory Tables; \texttt{sec\_not\_present} zeros out those entries,
thereby marking those pages as not present.  The memory region to be
protected is indicated via parameters passed to
\texttt{create\_nested\_pt} giving the start of the memory region to
be protected and its size.

Let \texttt{S0} be the initial state before \texttt{create\_nested\_pt} is
run.  Let \texttt{S1} be the state after it has run to termination, the
\texttt{CR3} register point to the top of the nested page tables, and
assume paging has been turned on.  Our final top-level theorems
state that the nested page tables set up by MinVisor are indeed the
desired page tables.

More specifically, we prove the following two facts:
\begin{itemize}
  \item If the precondition holds for \texttt{S0} and \texttt{addr} is
    disjoint from the memory region to be protected, then \\
    \texttt{(va-to-pa addr S1)} returns \texttt{addr}.
  \item If the precondition holds for \texttt{S0} and \texttt{addr} is
    not disjoint from the memory region to be protected, then 
    \texttt{(va-to-pa addr S1)} signals a page fault.
\end{itemize}

The precondition states that:
\begin{enumerate}
\item The initial Y86++ state is well formed;
\item The code is loaded at a specified location in memory;
\item The machine is poised to execute \texttt{create\_nested\_pt}
\item Paging is turned off on the machine;
\item The code, stack, and the tables being created are located at disjoint
  memory locations;
\item The tables are aligned on 4K page boundaries
\item The start of the memory region to be protected is aligned on a
  2MiB page boundary;
\item The size of the memory region to be protected is a non-zero
  multiple of 2MiB;
\item The stack does not wrap around memory;
\item All memory addresses used fit into a 32-bit register.
\end{enumerate}

\subsection{The ACL2 Model}
\label{A Sample Theorem}

In this section, we describe the modeling in ACL2 of the theorems
proved about \texttt{init\_pdts}.  This function initializes the four
page directory tables, and our theorem specifies its effects.  It is
formalized using the cutpoints method described above.

The C code for the proved functionality is listed in Appendix A.  This
is compiled using the \texttt{gcc} compiler into x86 assembler, which is hand
translated into Y86++ code.  The resulting Y86++ code for the
\texttt{init\_pdts} function is listed in Appendix B. This hand
translation could be automated, but we have not done so because we
hope to evolve our verification process to handle real x86 code.
However, as illustrated in section \ref{Comparison to the x86} above,
the translation is straightforward, so does not introduce significant
uncertainty into the overall process.  The Y86++ code is then
assembled into a Y86++ binary, and it is the binary that is the target
of verification.

The pre-condition of the execution (the 10 conditions listed in the
previous section) is modeled by an ACL2 predicate {\tt INIT\_PDTS-PRE}
on the initial state.  The results of the symbolic execution are
modeled by the ACL2 function {\tt INIT\_\-PDTS-\-MODIFY} specifying the
resulting state as a series of modifications to the initial state.
This specifies all changes to registers, flags, memory and other
system components.

This result function is tedious to construct and error-prone.  At some
point we hope to use ``wormhole abstraction''~\cite{HardinSmithYoung}
to elide state changes that are not of interest.  We believe this will
simplify these theorems considerably.  

Alternatively, with some modifications to the macros used to generate
the theorems, we could specify a series of ``read over write''
theorems as follows.  Let \texttt{S0} be the initial state before the
code being analyzed is run and \texttt{S1} the state after the code is
run to termination.  We can construct:
\begin{enumerate}
  \item A frame condition, specifying that reading in \texttt{S1} those
    portions of state not changed by the code, can be simplified to an
    equivalent read in \texttt{S0}.
  \item Theorems specifying, for those portions of state that are
    changed, how a read in \texttt{S1} would be reduced to reads in
    \texttt{S0}.
\end{enumerate}
This could be used to replace the effects function.  

The theorem we currently prove shows that running the machine on a
suitable initial state yields a resulting state identical to the
initial state except for specific concrete changes.

As with all ACL2 cutpoint proofs, the user supplies: the definition of
the language interpreter (in this case the Y86++ model), the
pre-conditions of the code's execution, and the expected modification
to the state.  Code involving loops also requires that appropriate
loop invariants be given.  A series of theorems are generated that,
if proven, are adequate to show that a terminating execution of code
from an initial state satisfying the precondition will produce the
indicated result.  Matthews, et al.~\cite{ACL2cutpoints} developed ACL2
macros that provide a convenient syntactic framework for supplying the
components and generating the requisite theorems.  See
\cite{ACL2cutpoints} for a thorough description of the methodology.

The following is the macro call that generates the forms
submitted to ACL2 in the proof of the function.
\begin{verbatim}
(defsimulate+
   y86-step
   :run y86
   :inmain in-init_pdts
   :cutpoint init_pdts-cutpoint-p
   :assertion-params (s0 s1)
   :precondition init_pdts-pre
   :assertion init_pdts-assertion
   :modify init_pdts-modify
   :exitsteps init_pdts-exitsteps
   :exists-next-exitpoint 
       init_pdts-exists-next-exitpoint
   :next-exitpoint init_pdts-next-exitpoint
   :correctness-theorem init_pdts-correct)
\end{verbatim}
Executing this form generates a complex ACL2 {\tt encapsulate} that
culminates in the following theorem: 
\begin{verbatim}
 (DEFTHM INIT_PDTS-CORRECT
   (IMPLIES 
      (AND (INIT_PDTS-PRE S1)
           (INIT_PDTS-EXISTS-NEXT-EXITPOINT S1))
      (AND (LET ((S1 (INIT_PDTS-NEXT-EXITPOINT S1)))
                (NOT (IN-INIT_PDTS S1)))
           (EQUAL (INIT_PDTS-NEXT-EXITPOINT S1)
                  (INIT_PDTS-MODIFY S1)))))
\end{verbatim}
This theorem is proved mechanically by ACL2.  Developing the proof,
however, did require the development of many subsidiary lemmas and
enhancements to the underlying libraries about our Y86++ model.
Specifically, we developed a library for specifying and reasoning
about disjoint memory regions.  We expect that this effort can be
amortized over many such proofs on the MinVisor project.

We have not yet considered termination in this proof, but believe that
that can be trivially added for the fixed finite structures we are
dealing with.

The cutpoints methodology has been used previously
\cite{ACL2cutpoints} in the verification of a JVM implementation of an
encryption/decryption system.  The MinVisor proofs may be the first
sizable application stressing this technology; we anticipate the need
to verify thousands of lines of binary code. 

\section{Related Work}
\label{Related Work}

Hypervisors were first introduced in the 1960s as a way to multiplex
scarce and expensive computing resources.  The advent of inexpensive
hardware and multitasking operating systems eroded their value in the
1980s and 1990s.  Over the last few years, however, hypervisors have
regained popularity as a versatile technology for enhancing security
and reliability.  One effect is that processor manufacturers are
removing obstacles that made it difficult to virtualize some system
resources on earlier processor designs without significant emulation
performance penalties.  Both Intel and AMD have developed
virtualization extensions to the x86 architecture~\cite{AMD,Intel}.

Xenon\cite{McDermott08,XenonPolicy} is a high assurance hypervisor
based on re-engineering Xen.  The designers have specified a formal
security policy based on the notion of
\textit{independence}~\cite{Roscoe}.  Though related to our work, the
Xenon hypervisor is much larger (around 70,000 lines of code) and the
effort is focused on gaining assurance through a policy-to-code
develop\-ment meth\-odology. In another large effort, Microsoft has developed
and used VCC~\cite{VCC}, A Verifier for Concurrent C, to verify components
of their 60,000 line Hyper-V hypervisor.

Several research efforts have demonstrated that it is possible to
construct small, robust and useful hypervisors.  The SecVisor
project~\cite{SecVisor} implemented two hypervisors (1739 and 1112
lines of code, respectively) supporting Linux kernel version 2.6.20.
Their systems provide strong integrity guarantees.  However, a
subsequent formal analysis~\cite{Franklin08} using the Mur$\phi{}$
model checker found two significant and exploitable design flaws.

Various projects have tied together hardware and software verification.  The
CLI Stack~\cite{CLIStack} was a collection of verified components
including a simple high-level language compiler, a linking loader, and
a microprocessor model.  The European Verisoft project~\cite{Verisoft}
has taken a similar approach with the goal of ``pervasive formal
verification'' of an entire computer system including hardware
(processor and devices), a real-time operating system, and
applications. This effort has since evolved into the Verisoft XT 
project and consortium.

The Fiasco project developed a microkernel running on x86 PCs and
intended to be compatible with the L4 microkernel.  The related
VFiasco effort attempted to prove security properties from the Fiasco
source code including a small hypervisor developed in C++.  A
``semantics compiler'' translates the C++ code into logical formulas
that are analyzed using PVS.  They also use a formal but incomplete
x86 model.~\cite{Hoffmann}

One component of the Robin Project~\cite{Nova} from Radboud University
Nijmegen in the Netherlands aims to verify the Nova hypervisor, a
micro-hypervisor targeted at the x86 architecture, using a very
similar approach to the VFiasco work.

A collaboration between NICTA, UNSW, Open Kernel Labs, and ANU resulted
in the creation and verification of seL4~\cite{seL4},
a third-generation L4 family microkernel. Their proof is done using
the Isabelle/HOL theorem prover, and involves proof of refinement
across three layers---an abstract specification, a functional
prototype in subset of Haskell, and an implementation in a subset of
C.

Other projects have attempted to apply formal methods to establish
certain separation properties.  Leslie, et al.~\cite{Leslie} have
developed a small microkernel/hypervisor in Haskell and specified a
combination of noninterference and information flow properties.  This,
they claim, provides a first step towards a formally verified
separation kernel, and reduces the gap between information flow theory
and operating system practice.  Alkassar and Paul~\cite{Alkassar}
report some initial work toward the verification of a ``baby''
hypervisor for the DLX RISC architecture.  A related effort is the
VeriOS project of the Saarland University (Germany).  Its goal is the
verification of an L4 kernel running on their own verified
microprocessor VAMP~\cite{VAMP}. The processor is modelled down to the
gate level using NuSMV in conjunction with Isabelle/HOL.

\section{Conclusions and Future Work}
\label{Conclusions and Future Work}

This project is a collaboration between the Systems and ACL2 groups at
the University of Texas at Austin.  As a result, our efforts have
been focused in two tracks:
\begin{enumerate}
\item Constructing the MinVisor hypervisor;
\item Developing the modeling and proof infrastructure to permit the
formal verification of MinVisor. 
\end{enumerate}
We have found this to be a highly useful collaboration in that
systems expertise was necessary to ensure that the artifact we built
is credible from a systems standpoint.  Our expertise in using ACL2
gave us hope that we could manage this complex verification
effort. 

Our goal has been to construct a simple but realistic hypervisor with
the goal of proving its correctness at the implementation level.
Though the effort is far from complete, we can report significant
progress, including:
\begin{itemize}
\item Building a simple, running hypervisor called MinVisor that
protects itself from the guest;
\item Enhancing the formal Y86 instruction set model to be able to
encode critical MinVisor functionality;
\item Developing proof techniques adequate to prove the correctness of
that functionality.
\end{itemize}

A project concern was to find a balance of a useful hypervisor and one
we could verify.  It is too early in the project to have definitive
evidence that our hypervisor is both credible and verifiable.  By
running on real hardware, we aimed to make the artifact interesting from
the systems perspective.  But our first cut is quite simple compared
to commercial hypervisors and even compared to SecVisor.  However, our
goal is to augment MinVisor with additional functionality once we have
developed sufficient proof infrastructure and facility doing the
proofs to convince ourselves that the project will succeed. 

On the systems side we hope to:
\begin{itemize}
\item Virtualize a small critical set of devices (e.g. console, timer, network)
     and block access to other devices.
\item Evolve MinVisor to support multiple guests.
\item Enhance CPU virtualization to multiplex among guests.
\item Enhance the fidelity with which the guest thinks it has real hardware.
\end{itemize}

On the verification track, we hope to do the following:
\begin{itemize}
\item Move from the Y86++ model we are currently using to an x86 model
as one becomes available. 

\item Prove rigorously the entire MinVisor code base. 

\item Make use of new ACL2 proof techniques to reason more efficiently
  about assembly code.
\end{itemize}

As we noted in Section 2, we hope to demonstrate that:
\begin{enumerate}
\item We can carry out the verification of a hypervisor at a very low
  level of abstraction to provide high assurance of correctness.
\item The resulting verified artifact is a credible piece of
  system software displaying useful hypervisor functionality.
\item The tools and techniques we develop will scale to apply to
  more realistic and full featured hypervisors. 
\end{enumerate}
Our progress to date makes us hopeful that we can achieve these
goals.  However, it is too early to claim success on any of them. 

\section*{Acknowledgements}
\label{Acknowledgements}

We thank our colleagues at the University of Texas, particularly
Sandip Ray, Warren Hunt, and J Moore for their work on the Y86, and
Matt Kaufmann for help with ACL2.  We'd particularly like to thank
Sandip Ray for helpful discussions throughout the project and comments
on the draft of this paper.  The comments from the anonymous reviewers
were also very helpful. This material is based upon work supported by
the National Science Foundation under Grant No.~CNS-0917162, and by
Raytheon under contract~200901296.

%\balancecolumns
\section*{Appendix}
\appendix
%Appendix A
\section{C Source Code}

We here give the C source for that part of MinVisor that sets up the
nested page tables.  We analyzed the corresponding Y86++ binary code.

\begin{verbatim}
typedef unsigned int u32;
typedef unsigned long long u64;

// pointer to the page-directory-pointer table
typedef u64 *pdpt_t;
// pointer to a page-directory table
typedef u64 *pdt_t;


void sec_not_present(pdpt_t pdptp, 
                     u32 *visor_start, 
                     u32 visor_size)
{
  u64 pdpt_entry;
  u64 tmp;
  u32 tmp32;
  pdt_t pdt;
  u32 start, end;
  u32 i, j;
  u64 mask;

  mask = ~((1 << 12) - 1);
  
  // The top two bits are the index into the 
  // 4 entry page-directory-pointer table 
  j = (u32)visor_start >> 30;
  pdpt_entry = pdptp[j];

  // mask off the lower 12 bits of pdpt_entry.
  tmp = pdpt_entry & mask;
  tmp32 = (u32)tmp;
  pdt = (pdt_t)tmp32; 

  // Bits 29-21 form the index into the 512 entry 
  // page-directory table.
  start = ((u32)visor_start & 0x3fe00000) >> 21;
  end = (((u32)visor_start + visor_size) 
                  & 0x3fe00000) >> 21;

  // mark not present from start to end
  for(i = start; i < end; i ++){
    pdt[i] = 0;
  }
}


void init_pdts(pdt_t pdt_array[4])
{
  u64 addr;
  pdt_t pdt;
  u32 i, j;
  u64 flags;
  u64 page_size_2m;

  // present, rw, user, accessed, dirty, 
  //   pse --- 2MiB page
  flags = 1 | 2 | 4 | 32 | 64 | 128;
  page_size_2m = 1 << 21;
  addr = 0;
  // 4 tables
  for(i = 0; i < 4; i ++){
    pdt = pdt_array[i];
    // 512 entries per table
    for(j = 0; j < 512; j ++)
    {
      // Make page directory entry for a 2MiB page.
      pdt[j] = addr | flags;
      addr += page_size_2m;
    }
  }
}


void init_pdpt(pdpt_t pdptp, pdt_t pdt_array[4])
{
  u32 i;
  u64 page_present;

  page_present = 1;
  
  for(i = 0; i < 4; i++){
    pdptp[i] = (u64)((u32)pdt_array[i] 
                           | page_present);
  }
}


pdpt_t create_nested_pt(pdpt_t pdptp, 
                        pdt_t pdt_array[4],
                        u32 *visor_start, 
                        u32 visor_size)

{
  init_pdpt(pdptp, pdt_array);

  init_pdts(pdt_array);

  sec_not_present(pdptp, visor_start, visor_size);

  return pdptp;
}
\end{verbatim}

% This next section command marks the start of
% Appendix B, and does not continue the present hierarchy
\section{Y86++ Source Code for init\_pdts}

We here give the Y86++ assembly corresponding to one of the above
functions. \texttt{init\_pdts}.

\begin{verbatim}
    (:init_pdts
    (pushl :ebp)
    (rrmovl :esp :ebp)
    (pushl :esi)
    (pushl :ebx)
    (irmovl 48 :imme1)
    (subl :imme1 :esp)
    (irmovl 231 :imme1)
    (rmmovl :imme1 -24 (:ebp))
    (irmovl 0 :imme1)
    (rmmovl :imme1 -20 (:ebp))
    (irmovl 2097152 :imme1)
    (rmmovl :imme1 -16 (:ebp))
    (irmovl 0 :imme1)
    (rmmovl :imme1 -12 (:ebp))
    (irmovl 0 :imme1)
    (rmmovl :imme1 -48 (:ebp))
    (irmovl 0 :imme1)
    (rmmovl :imme1 -44 (:ebp))
    (irmovl 0 :imme1)
    (rmmovl :imme1 -32 (:ebp))
    (jmp :L7)
    :L8
    (mrmovl -32 (:ebp) :eax)
    (irmovl 2 :imme1)
    (sall :imme1 :eax)
    (mrmovl 8 (:ebp) :valu1)
    (addl :valu1 :eax)
    (mrmovl 0 (:eax) :eax)
    (rmmovl :eax -36 (:ebp))
    (irmovl 0 :imme1)
    (rmmovl :imme1 -28 (:ebp))
    (jmp :L9)
    :L10
    (mrmovl -28 (:ebp) :eax)
    (irmovl 3 :imme1)
    (sall :imme1 :eax)
    (rrmovl :eax :esi)
    (mrmovl -36 (:ebp) :valu1)
    (addl :valu1 :esi)
    (mrmovl -24 (:ebp) :ecx)
    (mrmovl -20 (:ebp) :ebx)
    (mrmovl -48 (:ebp) :eax)
    (orl :ecx :eax)
    (mrmovl -44 (:ebp) :edx)
    (orl :ebx :edx)
    (rmmovl :eax 0 (:esi))
    (rmmovl :edx 4 (:esi))
    (mrmovl -16 (:ebp) :eax)
    (mrmovl -12 (:ebp) :edx)
    (mrmovl -48 (:ebp) :valu1)
    (addl :eax :valu1)
    (rmmovl :valu1 -48 (:ebp))
    (mrmovl -44 (:ebp) :valu1)
    (adcl :edx :valu1)
    (rmmovl :valu1 -44 (:ebp))
    (irmovl 1 :imme1)
    (mrmovl -28 (:ebp) :valu1)
    (addl :imme1 :valu1)
    (rmmovl :valu1 -28 (:ebp))
    :L9
    (irmovl 511 :imme1)
    (mrmovl -28 (:ebp) :valu1)
    (cmpl :imme1 :valu1)
    (jbe :L10)
    (irmovl 1 :imme1)
    (mrmovl -32 (:ebp) :valu1)
    (addl :imme1 :valu1)
    (rmmovl :valu1 -32 (:ebp))
    :L7
    (irmovl 3 :imme1)
    (mrmovl -32 (:ebp) :valu1)
    (cmpl :imme1 :valu1)
    (jbe :L8)
    (irmovl 48 :valu1)
    (addl :valu1 :esp)
    (popl :ebx)
    (popl :esi)
    (popl :ebp)
    (ret)))
\end{verbatim}

%\bibliographystyle{abbrv}
\bibliography{paper}  % paper.bib is the name of the Bibliography in this case
\bibliographystyle{eptcs}
% ACM needs 'a single self-contained file'!
%

% \section{References}
% Generated by bibtex from your ~.bib file.  Run latex,
% then bibtex, then latex twice (to resolve references)
% to create the ~.bbl file.  Insert that ~.bbl file into
% the .tex source file and comment out
% the command \texttt{{\char'134}thebibliography}.

%\balancecolumns
% That's all folks!
\end{document}